\documentclass[a4pape,citesort]{PoS}
\usepackage{caption}
\usepackage{subcaption}
\usepackage{graphicx}
\usepackage{citesort}
\usepackage{bold-extra}

\def\ncteqpp{{\tt \textbf{nCTEQ++}}}
\def\wz{$W^\pm\!/Z$}

\usepackage{wrapfig}
\usepackage{lipsum,lineno}
\usepackage{microtype}  

\makeatletter 
\renewcommand\speaker[1]{\if@speaker\global\@dblspeaktrue\fi
			\global\@speakertrue
			\global\setbox\@firstaubox
			\hbox{{\let\thanks\@gobble
				\let\footnote\@gobble\small
				\rm  The nCTEQ Collaboration}}%
			#1\thanks{Speaker.}\
			}%
\makeatother

\title{PDF Flavor Determination and the nCTEQ PDFs: 
 \hspace{\textwidth} 
{\it  \wz{}  vector boson production in heavy ion collisions}
}

\ShortTitle{PDF Flavor Determination and the nCTEQ PDFs}

\author{%
The  nCTEQ Collaboration:\thanks{%
We acknowledge the hospitality of CERN, DESY, and Fermilab where a
portion of this work was performed.
This work was also partially supported by the U.S.\ Department of
Energy under Grant No.\ DE-SC0010129.
}\qquad  \null  \hspace{\textwidth}
E.~Godat\rlap,${}^1$ 
D.~B.~Clark\rlap,${}^1$ 
T.~J.~Hobbs\rlap,${}^1$ 
T.~Je\v{z}o\rlap,${}^2$ 
J.~Kent\rlap,${}^1$ 
C.~Keppel\rlap,${}^3  $
K.~Kova\v{r}\'{i}k\rlap,${}^4$ 
A.~Kusina\rlap,${}^{5,6}$ 
F.~Lyonnet\rlap,${}^1$ 
J.G.~Morfin\rlap,${}^7$
F.~I.~Olness\rlap,${}^1$\speaker{}
J.F.~Owens\rlap,${}^8$
I.~Schienbein\rlap,${}^5$ 
J.~Y.~Yu${}^1$ 
\\
${}^1$Southern Methodist University, Dallas, TX 75275, USA\\ 
${}^2$Physik-Institut, Universit\"at Z\"urich, Winterthurerstrasse 190, CH-8057 Z\"urich,
Switzerland\\
${}^3$Thomas Jefferson National Accelerator Facility, Newport News, VA, 23606, USA\\
${}^4$Institut f\"{u}r Theoretische Physik, Westf\"{a}lische Wilhelms-Universit\"{a}t
M\"{u}nster, \\ \qquad
Wilhelm-Klemm-Stra{\ss}e 9, D-48149 M\"{u}nster, Germany \\
${}^5$Laboratoire de Physique Subatomique et de Cosmologie, Universit\'{e}
Grenoble-Alpes, \\ \qquad
CNRS/IN2P3, 53 avenue des Martyrs, 38026 Grenoble,
France \\
${}6$Institute of Nuclear Physics, Polish Academy of Sciences, \\ \qquad
ul. Radzikowskiego 152, 31-342 Cracow, Poland\\
${}^7$Fermi National Accelerator Laboratory, Batavia, Illinois 60510, USA\\
${}^8$Department of Physics, Florida State University, Tallahassee, Florida 32306-4350, USA\\
}

\abstract{

Recent LHC \wz{}  vector boson production
data in proton-lead  collisions are quite sensitive to the
heavier flavors (especially the strange PDF), and this complements the
information from neutrino-DIS data. As the proton flavor determination
is dependent on nuclear corrections (from heavy target DIS, for
example), LHC heavy ion measurements can also help improve proton PDFs.
We introduce a new  implementation of the nCTEQ code (\ncteqpp)
based on C++  which has a modular strucure and enables us to 
easily integrate programs such as HOPPET, APPLgrid, and MCFM.
Using ApplGrids generated from MCFM, 
we use \ncteqpp{} to perform a  fit including the $pPb$ LHC \wz{}  vector boson  data. 
}

\FullConference{XXVI International Workshop on Deep-Inelastic Scattering and Related Subjects (DIS2018)\\
		16-20 April 2018\\
		Kobe, Japan}

\begin{document}

\def\figscheme{
\begin{figure}[t] 
\centering{} 
\includegraphics[width=0.95\textwidth]{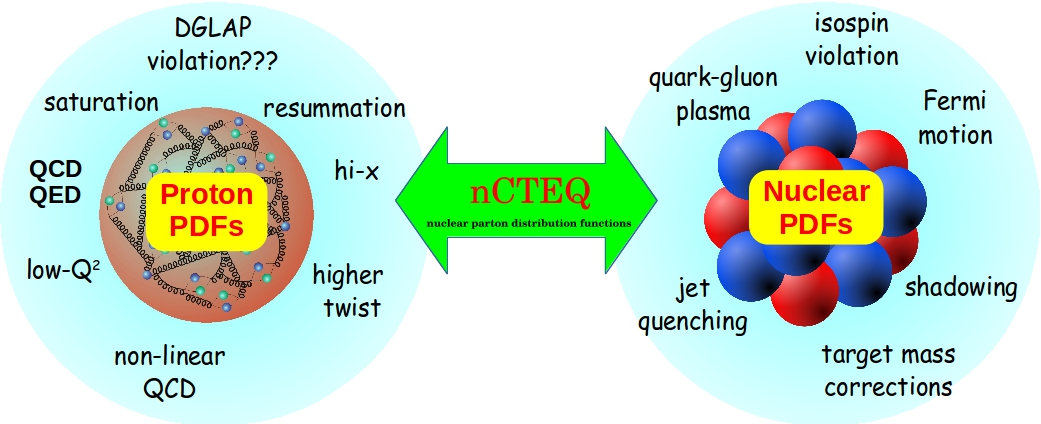}
\vspace{0pt}
\caption{Schematic representation of selected 
phenomenological issues 
that can impact the determination of proton and nuclear PDFs.\cite{Kovarik:2015cma}
}
\label{fig:scheme}
\end{figure} 
}

\def\figatlas{
\begin{wrapfigure}{R}{0.450\textwidth} 
\centering{} 
\vspace{-35pt}
\includegraphics[width=0.450\textwidth]{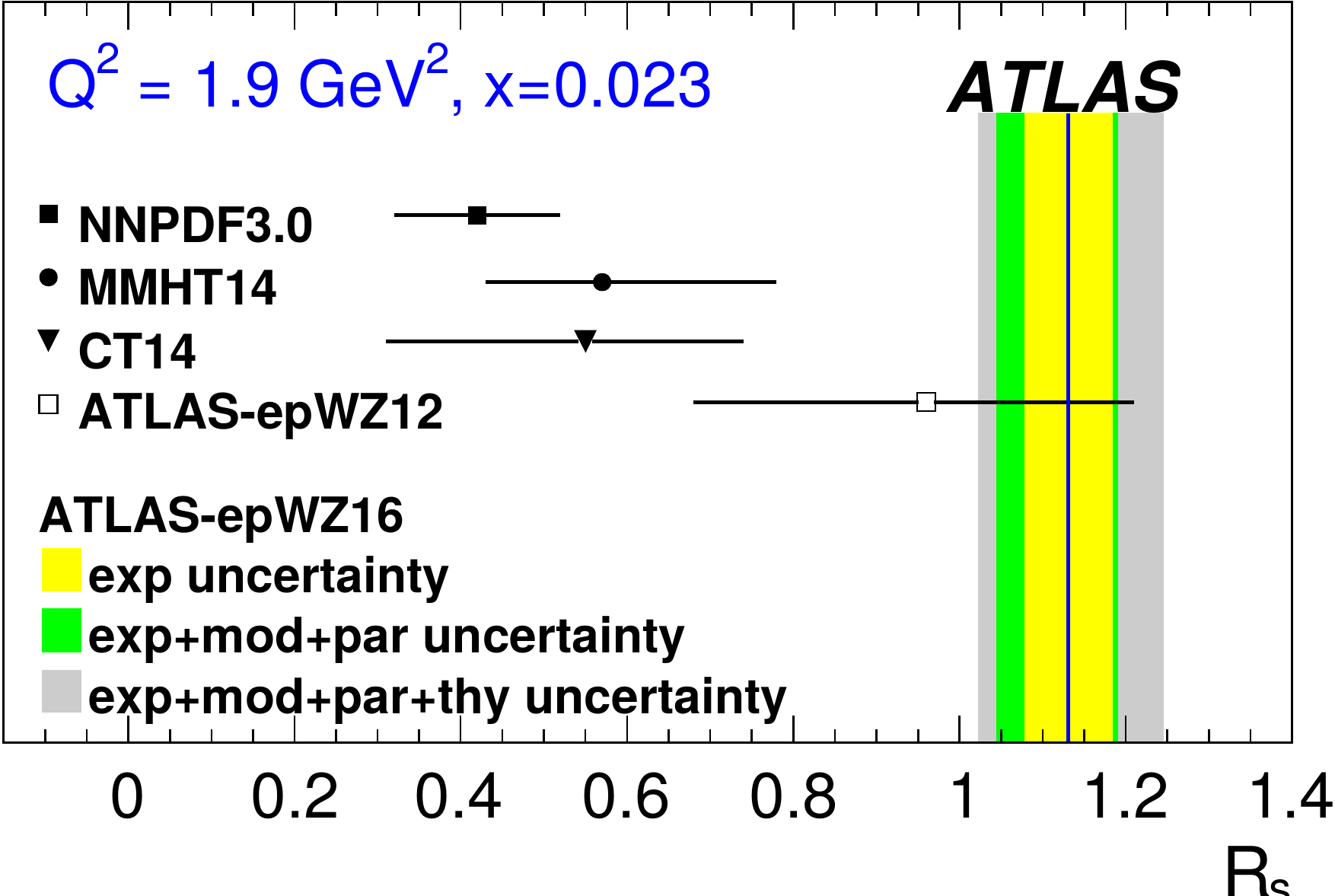}
\vspace{-10pt}
\caption{
The relative quark ratio $R_s={(s+\bar{s})/(\bar{u}+\bar{d})}$ 
as measured by the ATLAS collaboration from  $W/Z$  
production in proton-proton collisions~\cite{Aaboud:2016btc}.
}
\label{fig:atlas}
\end{wrapfigure}
}

\def\figchi{
\begin{wrapfigure}{R}{0.65\textwidth} 
\centering{} 
\vspace{-25pt}
\includegraphics[width=0.65\textwidth]{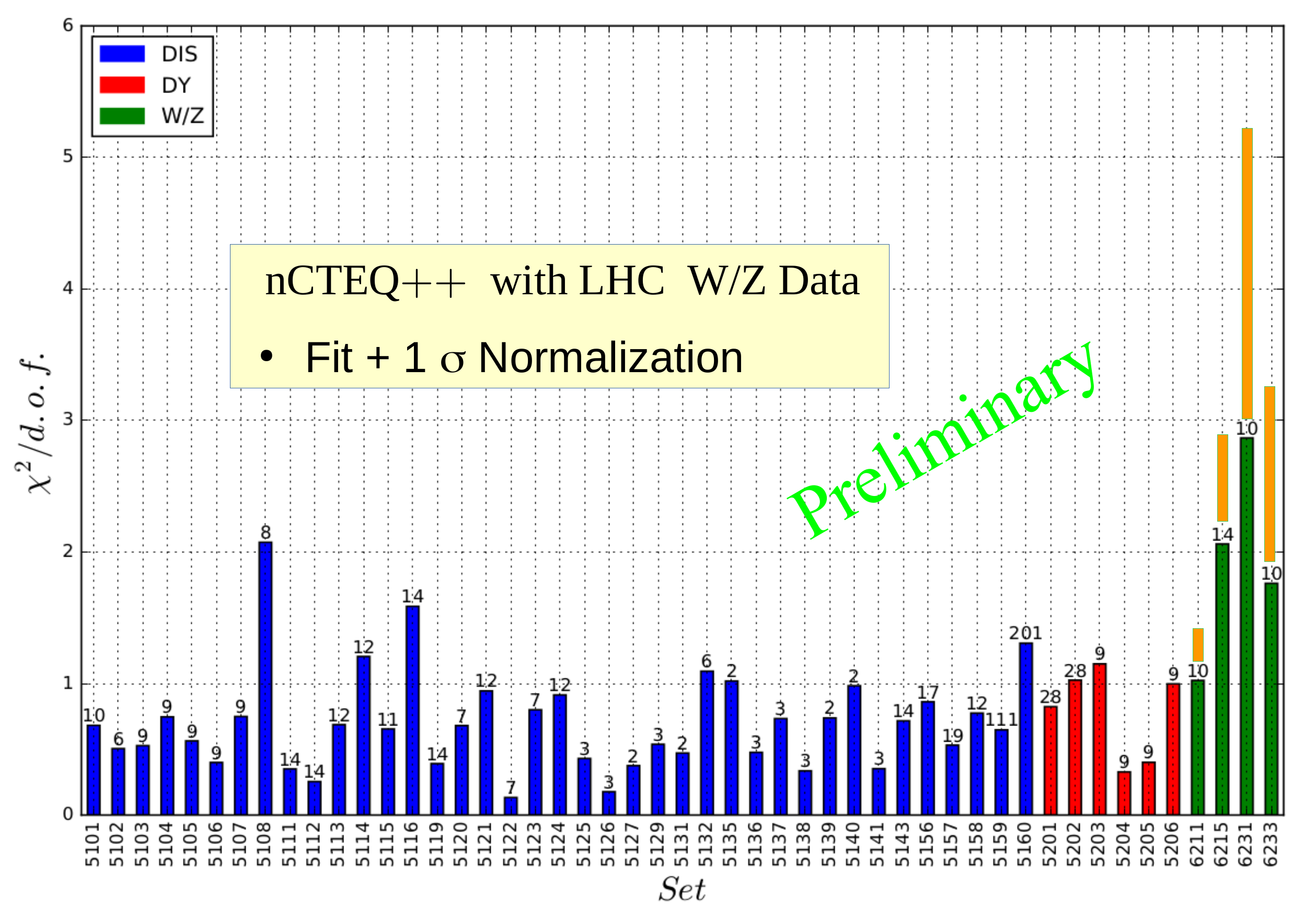}
\vspace{-20pt}
\caption{
$\chi^2/d.o.f$  for each data set included in nCTEQ+LHC. The individual
data sets are identified by the ID number corresponding to those in Ref.~\cite{Kusina:2016fxy}.
The  LHC \wz{} data is displayed in green and have ID numbers corresponding to 62XX;
in this fit, we have allowed a floating normalization of $1\, \sigma$.
}
\label{fig:chi2}
\end{wrapfigure}
}

\def\figcorr{
\begin{wrapfigure}{R}{0.60\textwidth} 
\centering{} 
\vspace{-25pt}
\includegraphics[width=0.60\textwidth]{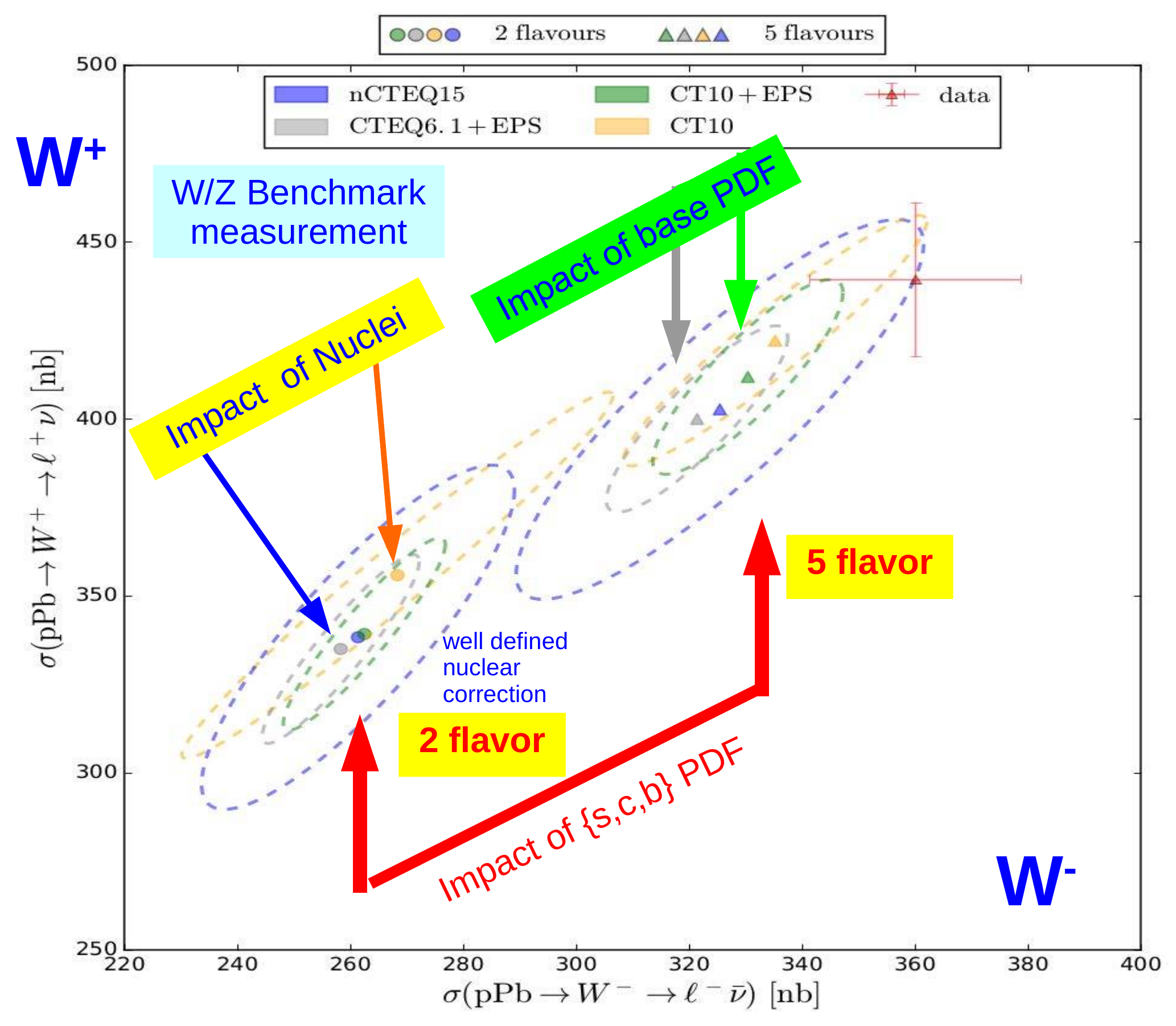}
\vspace{-20pt}
\caption{
Correlations between $W^+$ and $W^-$ cross sections calculated with different input PDFs and assumptions
for the $pPb$ process. 
The upper-right ellipse is computed with all 5 flavors, and the lower-left ellipse  
includes only the $\{u,d \}$ partons. 
 We show here results for
nCTEQ15, EPS09+CT10, EPS09+CTEQ6.1 and CT10
PDFs 
with the CMS data.
}
\label{fig:corr}
\end{wrapfigure}
}

\def\figdy{
\begin{wrapfigure}{R}{0.65\textwidth} 
\centering{} 
\vspace{-25pt}
\includegraphics[width=0.65\textwidth]{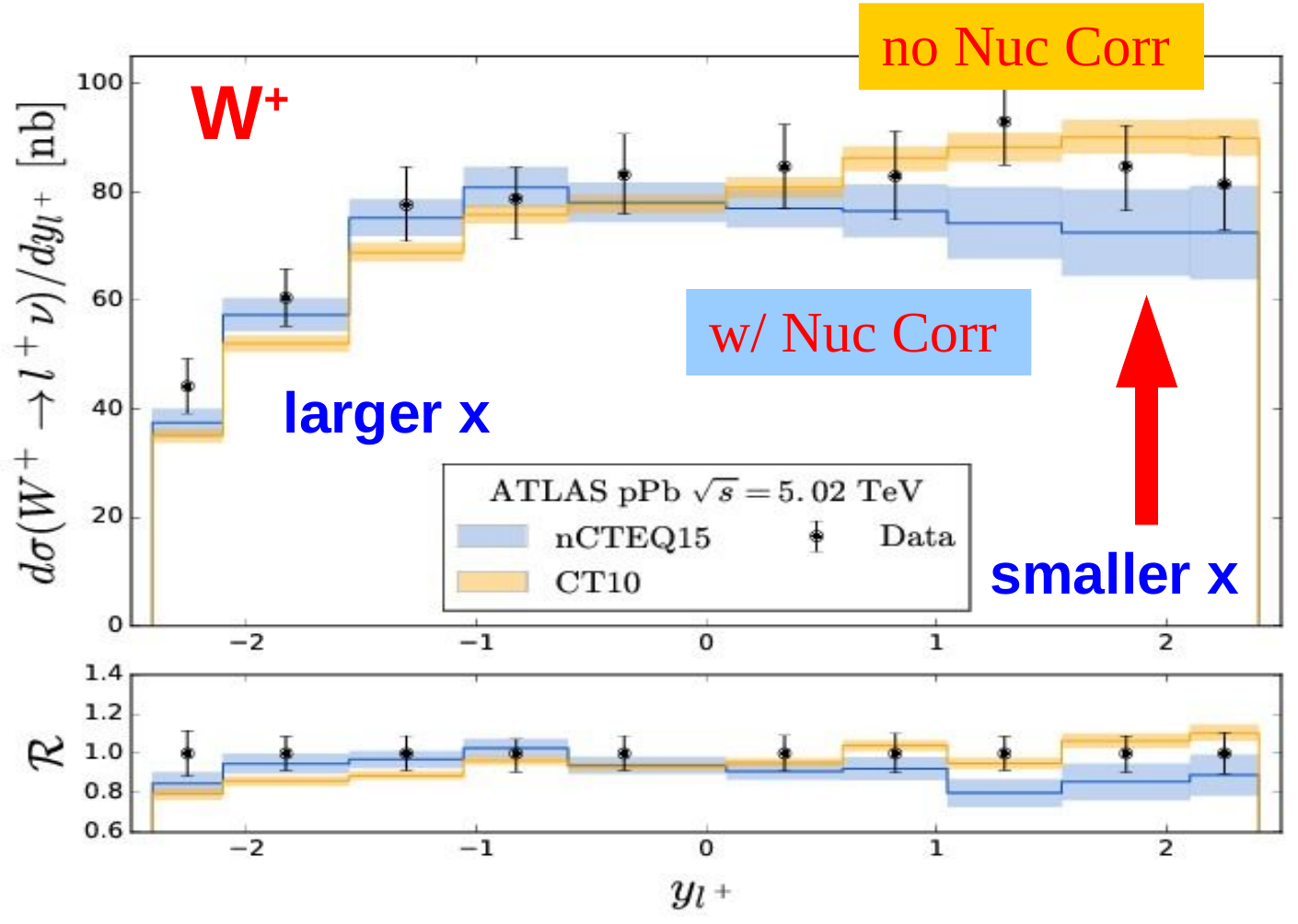}
\vspace{-20pt}
\caption{
corr fig
}
\label{fig:dsigdy}
\end{wrapfigure}
}

\def\figdouble{
\begin{figure}[t] 
\centering{} 
\includegraphics[width=0.45\textwidth]{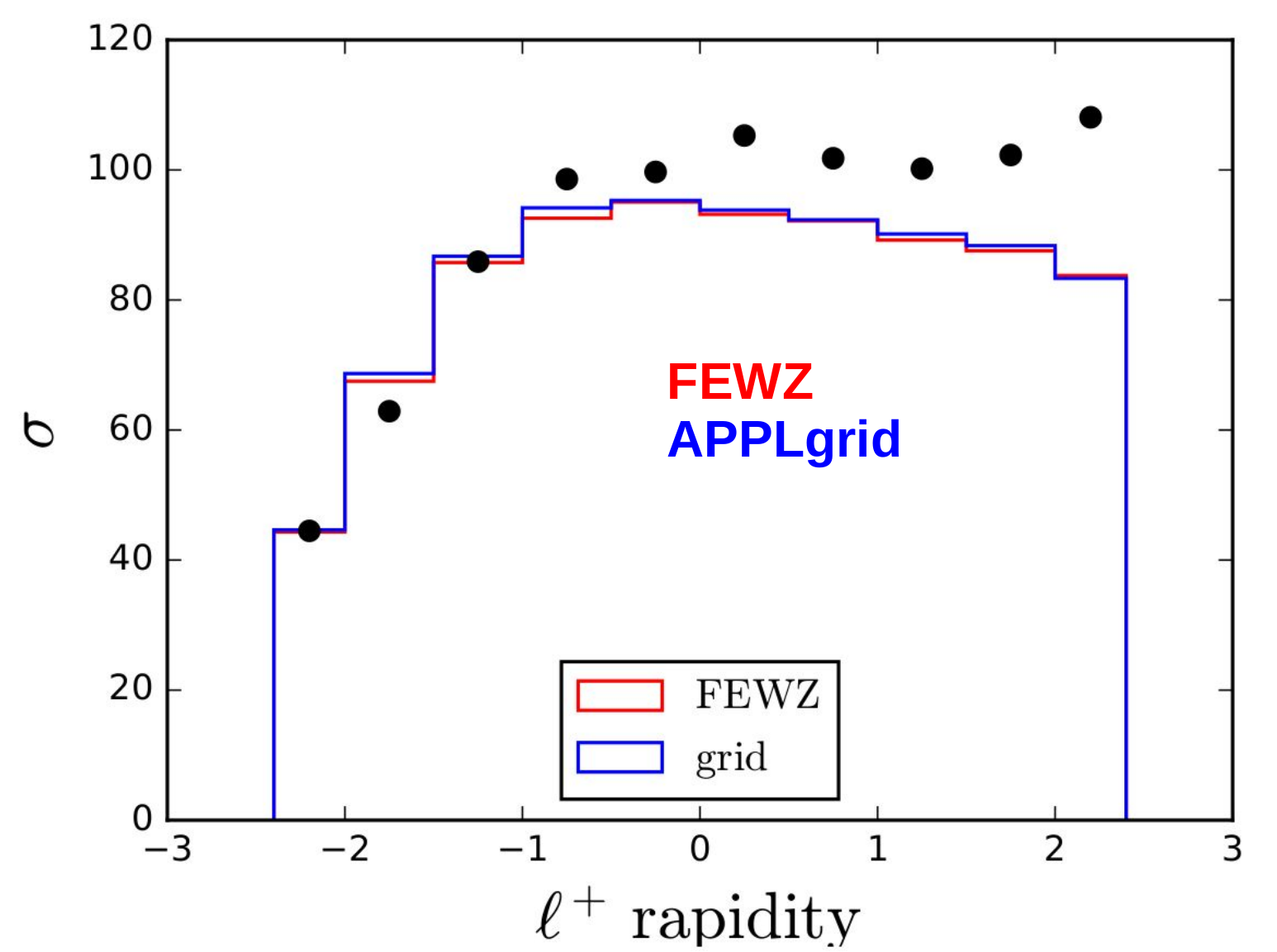}
\hfil
\includegraphics[width=0.45\textwidth]{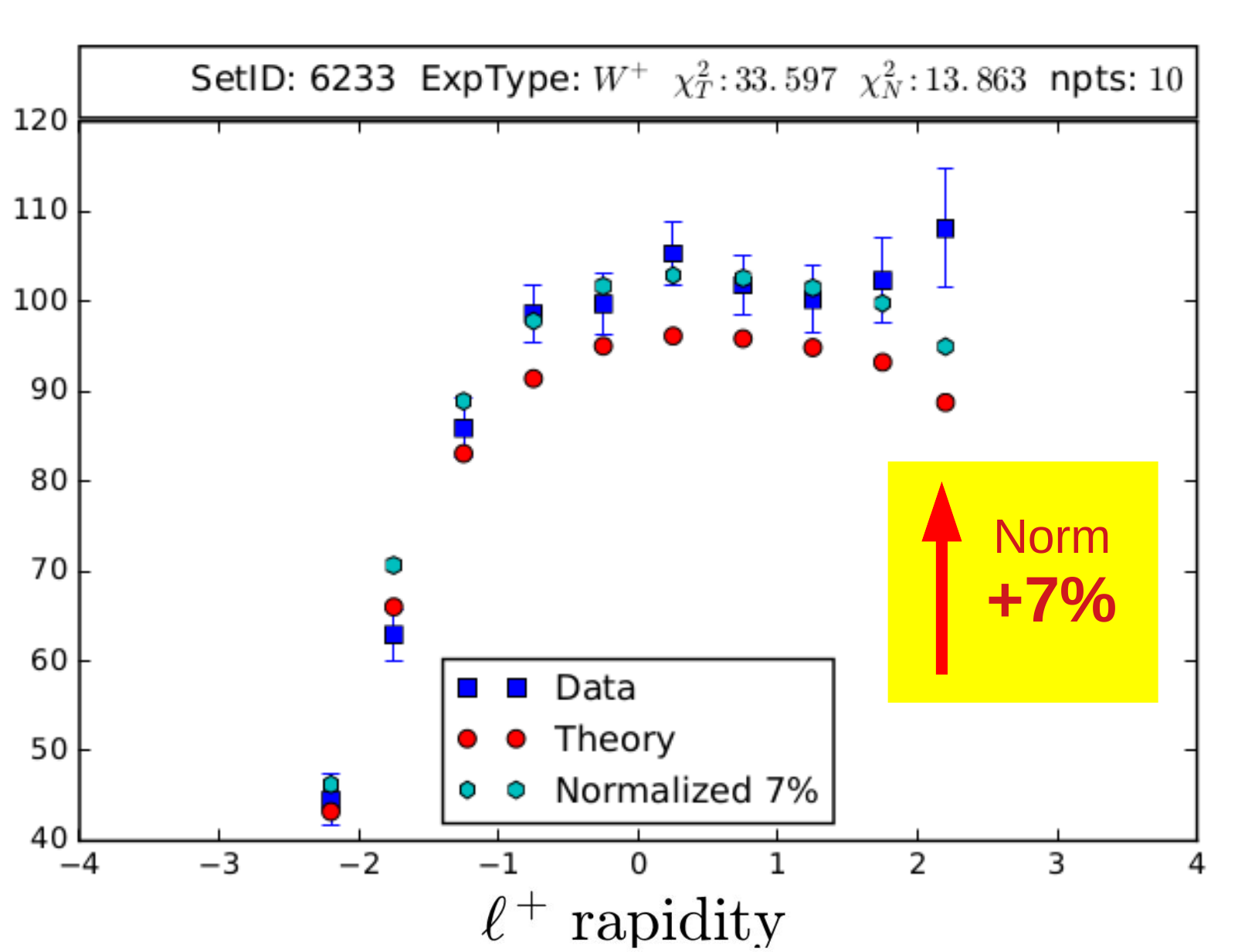}
\vspace{0pt}
\caption{
a)~Theory predictions for $W^+$ production in $pPb$  from FEWZ (red),
an APPLgrid  calculation (blue),
and the associated CMS data set (black points).
The APPLgrids were generated using replica grids from MCFM in $pp$ mode, 
and then applied to the  $pPb$ calculation. 
%
\quad
b)~A Sample comparison of nCTEQ+LHC fit  to the CMS data 
 as well as a (2$\sigma$) normalized theory prediction.
}
\label{fig:fewz}
\end{figure} 
}

\null \vspace{-1.0cm}
\section{Introduction}
\nocite{Kovarik:2015cma}
\null \vspace{-0.5cm}

\figscheme
\figatlas{}

The Parton Distribution Functions (PDFs) are the key ingredient that
enables us to connect experiment with theory using the QCD
improved parton model to describe the distribution of quarks and
gluons in the proton.
Despite decades of studies, there is yet much to learn about the
proton structure.  

A very interesting  result which was discussed
at this meeting was the ratio of the strange PDF to the up- and
down-sea quark PDFs extracted using  \wz{}  production 
(a ``standard candle'' measurement)
at the LHC 
as shown in Figure~\ref{fig:atlas}.
This case is just one example where improved determinations of
the PDFs can significantly impact precision theoretical  predictions,
and thus enhance our ability to discern ``new physics'' signatures
from uncertain ``standard model'' processes.
The goal of the nCTEQ collaboration is to make maximal use of the
available data, both proton and nuclei, to obtain the most precise
determination of the PDFs. In this brief report, we summarize some of
the recent advancements toward this goal.

\null \vspace{-1.0cm}
\section{The nCTEQ Project}
\null \vspace{-0.5cm}

The  nCTEQ project\footnote{For details, see {\tt www.ncteq.org} which is hosted at HepForge.org.} 
is built upon the work of the CTEQ proton PDF global fitting effort by 
extending the fit degrees of freedom into the nuclear dimension. 
Previous to the nCTEQ effort, nuclear data was ``corrected'' to isoscalar data 
and added to the proton PDF fit {\it without} any uncertainties. 
In contrast, the nCTEQ framework allows full communication between the nuclear data 
and the proton data, as illustrated schematically in Fig.~\ref{fig:scheme}.
This enables us to investigate if observed tensions between
data sets could potentially be  attributed to the nuclear corrections.

The details of the nCTEQ program are presented in Ref.~\cite{Kovarik:2015cma}.
The analysis includes   Deeply Inelastic Scattering (DIS), 
lepton pair production (Drell-Yan), and pion production 
from a variety of experiments totaling 740  data (after cuts) and 19 nuclei.
The  computed PDFs compare favorably to other determinations from the 
literature~\cite{Hirai:2007sx,deFlorian:2011fp,Eskola:2016oht}.

\section{LHC Heavy Ion $\boldmath{W}$ Production and Correlations}
\figcorr{}

In this report, we will focus on \wz{}  production as 
this process is sensitive to the heavier  flavors. 
In Fig.~\ref{fig:corr} we display the correlations between $W^+$ and $W^-$ 
cross sections for proton-lead interactions calculated with different input PDFs and assumptions~\cite{Kusina:2016fxy}.
By comparing the results with and without the $\{s,c,b\}$ flavors, we see the heavier quarks do have a
large impact on this observable;  hence, this process can provide incisive information about the 
corresponding PDFs.

To see the effect of the nuclear corrections, we can compare the CT10 proton result with the other calculations. 
We observe that for the case of 2 flavors only, the separation of the proton result  (CT10) and the 
nuclear results are quite distinct. In this case, the effect of the specific nuclear correction (nCTEQ15 or EPS09)
or the effect of the underling base PDF (CTEQ6.1 or CT10) is minimal. 

In contrast, when we compare this picture to the 5 flavor results, the division between the proton and nuclear
result is not as simple as the different nuclear corrections and base PDFs yield a broader range of results. 
In particular, we note that the strange quark PDFs in the CTEQ6.1 and CT10 base PDFs are quite different, and this
will contribute to the spread of results. 

Thus, proton-lead  production of \wz{}  is an ideal ``laboratory'' 
as this process is sensitive to i)~the heavy flavor components,
ii)~the nuclear corrections, and iii)~the underlying ``base'' PDF.

\newpage
\section{The \ncteqpp{} Program \&  Fast Grid Computations}
\figdouble{}

The original nCTEQ project was based on the {Fortran\,77} proton PDF framework.\cite{Schienbein:2007fs}
As the scope of this project has grown, it was necessary to restructure 
the nCTEQ code to make it more modular. 
The new \ncteqpp{} program is C++ running on top of legacy Fortran; 
the evolution is performed by a modified version of HOPPET\cite{Salam:2008qg} 
(extended to accommodate grids of multiple nuclei), and the output of the fit 
is exported in YAML format  and then processed by Python Jupyter notebooks.

If we are to include complex NLO processes into our PDF fit in an efficient manner,
it is essential to make use of the grid tools such as APPLgrid,\cite{Carli:2010rw} and we have implemented 
these techniques into our new \ncteqpp{} framework. 
Specifically, \ncteqpp{}  uses pre-computed APPLgrid grids generated by
the MCFM {{Fortran}}   program\cite{Campbell:2015qma} to perform fast NLO and NNLO \wz{} calculations 
inside the Minuit fitting loop.
We demonstrate the advantages of these features below. 

In 
Figure~\ref{fig:fewz}-a) we validate our grid implementation 
for nuclear \wz{}  production. 
The FEWZ cross section\cite{Li:2012wna} (red histogram) was computed using a modified version of the standard FEWZ 
program to accommodate the proton-lead initial state.\cite{Kusina:2016fxy}
 Although  such modifications are technically
 straightforward,  their systematic  implementation and cross-checking is time consuming nevertheless.
In contrast for the APPLgrid calculation (blue histogram), the MCFM program was used (in proton-proton mode) 
to generate a collection of grids. If sufficient statistics are  used, the combined grids are 
independent of the initial PDF. 
\begin{quote}
Thus, we can generate grids in proton-proton mode, but then 
use them also for \hbox{proton-nucleus} or nucleus-nucleus processes! 
\end{quote}
\noindent
The fact that the red and blue histograms
match within statistical accuracy validates this approach. 

Moreover, now that we have implemented this APPLgrid framework, we have access to all the $\sim\!\!1000$ processes 
included in
  MCFM.\footnote{%
  For example, a recent study~\cite{Kusina:2017gkz} uses  heavy flavor meson
data to constrain the  gluon distribution; this 
data can also can be included  using APPLgrid techniques.
} 
We now illustrate the utility of this flexible modular framework by performing a
(preliminary) fit including the LHC \wz{} heavy ion data.

\section{PDF fit to LHC \boldmath{\wz{}} Data }
\nobreak
\figchi{}
\nobreak

Now that we can compute  fast NLO (or NNLO)  \wz{}  cross sections, 
we can include these directly into our PDF fits.
A sample comparison is  displayed in 
Figure~\ref{fig:fewz}-b).
The data for CMS $W^+ $are shown as blue squares with error bars, and the theory in the red circles.
For comparison, we also present the theory with a 7\% normalization shift applied; the quoted 
luminosity uncertainty is 3.5\%.

In Figure~\ref{fig:chi2} we show the computed $\chi^2$ 
results for the individual experiments before and after our fit.
As we are only opening up a limited number of parameters in this preliminary fit, it is primarily 
the LHC \wz{}  data that is affected.\footnote{%
  Specifically, we fit 12 parameters: 3 for  $ s+\bar{s} $,
  and the remaining 9 for $\{g, u_V, d_V, \bar{u}+\bar{d} \}$.
This is in contrast to  nCTEQ15 which fits 16 parameters  for $\{g, u_V, d_V, \bar{u}+\bar{d} \}$ and keeps the strange PDF fixed.
}
The separate processes in the figure are color coded.
The DIS data is represented by blue bars and
the Drell-Yan data by red bars;  $\chi^2$ of these data sets is essentially unchanged.
The  \wz{}  is  represented by the green bars, and the orange bars 
show the change in $\chi^2$ before and after the fit.
We have allowed a variable normalization of the data (with an appropriate $\chi^2$ penalty);
our preliminary results indicated a normalization shift of $1 \sigma$ (2.7\% for ATLAS and 3.5\% for CMS)
yield near optimal values.\footnote{%
  In Fig.~\ref{fig:fewz}-b) we display a normalization shift of $2 \sigma$ for illustration purposes; however,
  when the normalization penalty is included, the  $1 \sigma$ shift yields a lower total $\chi^2$.
  }
We see that by including
the  \wz{}  data into the fit we are able to obtain a much improved description of this data set. 
A full fit (with a full set of free parameters) is underway,  but this preliminary fit 
with a limited set of free parameters is sufficient to demonstrate the merits 
the new \ncteqpp{} code with the grid-based computations.

\newpage
\section{Conclusion}
\null \vspace{-0.5cm}

The goal of the nCTEQ project is to obtain the most precise PDFs using 
the full collection of both proton and nuclear data. 
In this brief report we have observed that the LHC heavy ion data can help
determine nuclear corrections for large  $A$ values in a 
kinematic $\{x,Q^2\}$ range very different from those 
provided by fixed-target measurements.

The \wz{} \ \     pPb data are particularly sensitive to the heavier quark flavors
(especially the strange PDF), so this provides important information
on the  flavor decomposition.
The new \ncteqpp{} framework, which integrates the 
NLO grid calculations, allows us to easily include these  processes into 
the PDF fit. 

Improved information on the nuclear corrections from the LHC lead data
can also help reduce proton PDF uncertainties as (at present) fixed-target nuclear
data is essential for distinguishing the individual flavors~\cite{Ball:2017nwa}.
The next step is to extend the above preliminary fit with a complete set 
of free parameters and additional data sets  to
help separately disentangle issues of flavor differentiation and nuclear corrections. 

 \vspace{-0.0cm}


\end{document}